\documentclass[aps,prl,reprint,superscriptaddress,footinbib]{revtex4-1}

\usepackage{amssymb}
\usepackage{bm}
\usepackage{graphicx}
\usepackage{amsmath}

\begin{document}


\title{ Transient response in granular quasi-2D bounded heap flow}


\author{Hongyi Xiao}
\affiliation{Department of Mechanical Engineering, Northwestern University, Evanston, Illinois 60208, USA}

\author{Julio M. Ottino}
\affiliation{Department of Mechanical Engineering, Northwestern University, Evanston, Illinois 60208, USA}
\affiliation{Department of Chemical and Biological Engineering, Northwestern University, Evanston, Illinois 60208, USA}
\affiliation{The Northwestern Institute on Complex Systems (NICO), Northwestern University, Evanston, Illinois 60208, USA}

\author{Richard M. Lueptow}
\email{r-lueptow@northwestern.edu}
\affiliation{Department of Mechanical Engineering, Northwestern University, Evanston, Illinois 60208, USA}
\affiliation{The Northwestern Institute on Complex Systems (NICO), Northwestern University, Evanston, Illinois 60208, USA}

\author{Paul B. Umbanhowar}
\affiliation{Department of Mechanical Engineering, Northwestern University, Evanston, Illinois 60208, USA}


\date{\today}

\begin{abstract}
We study the transition between steady flows of non-cohesive granular materials in quasi-2D bounded heaps by suddenly changing the feed rate. In both experiments and simulations, the primary feature of the transition is a wedge of flowing particles that propagates downstream over the rising free surface with a wedge front velocity inversely proportional to the square root of time. An additional longer duration transient process continues after the wedge front reaches the downstream wall. The entire transition is well modeled as a moving boundary problem with a diffusion-like equation derived from local mass balance and a local relation between the flux and the surface slope.
\end{abstract}


\maketitle

\par Heaps of granular materials form in both geophysical and industrial systems, and exhibit kinematics that vary in both the streamwise and depthwise directions~\cite{Khakhar2001,Fan2013}. While most previous studies of quasi-2D bounded heap formation have considered steady feed rates exclusively, e.g.~\cite{grasselli1999shapes,Khakhar2001,Fan2012,Fan2013,fan2015shear,Xiao2016,Fan2017}, recent work with size bidisperse mixtures of spherical granular particles fed onto a quasi-2D bounded heap under alternating feed rates shows dramatic changes to the segregation pattern, indicating the existence of complex transient flows~\cite{Xiao2017}. This work raises the question of how granular flows relax to steady-state following a change to a control parameter such as the feed rate.

\par In bounded heap flows, particles travel down the surface in a thin flowing layer over a static bed with a rising free surface inclined at an angle determined by particle properties (e.g., shape and friction), the sidewall gap, and the feed rate~\cite{jop2005crucial,Fan2012}. The evolution of the heap is determined by the surface rise velocity and particle exchange between the flowing layer and the underlying static bed~\cite{Bouchaud1994,Boutreux1998,Khakhar2001,Douady2002}. For steady feed rates, the entire free surface rises at a constant velocity $v_r=q/W$, where $q$ is the volumetric feed rate divided by the gap thickness and $W$ is the heap width~\cite{Fan2013,Boutreux1998,Khakhar2001}. However, it is unclear how material deposition varies in unsteady processes.

\par In this Letter, we experimentally and computationally study the transient processes in a quasi-2D bounded heap during single transitions between steady states at different feed rates using monodisperse spherical particles. During transition we observe a developing wedge on the rising surface with a downstream propagating front. The wedge front arises from the change in the feed rate and differs from an avalanche front triggered by a sudden release of material that propagates with a constant velocity~\cite{daerr1999two,borzsonyi2005two,edwards2015erosion,saingier2016front}, and from transient phenomena in tumbler flows~\cite{CourrechDuPont2005,Pohlman2009} or streamwise invariant flows~\cite{Jop2007,parez2016unsteady,Capart2015,taberlet2004growth}. Instead, our experiments and Discrete Element Method (DEM) simulations show that the wedge front velocity is proportional to $t^{-1/2}$, and that the surface undergoes a slow relaxation that continues long after the wedge front reaches the downstream bounding wall. We show that these transient processes originate in the relation between the local surface slope and the local flow rate in depositing flows, which leads to a model with the same form as the diffusion equation that accurately predicts the observed dynamics.


  \begin{figure*}
  	\centerline{\includegraphics[width=7.2 in]{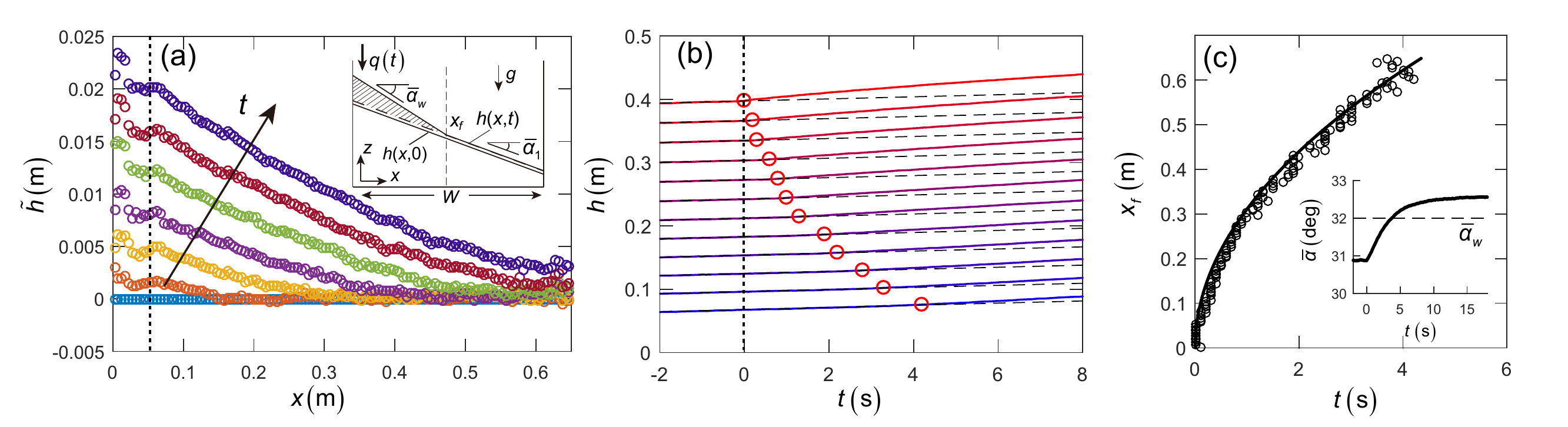}}
  	\vspace*{-10 pt}
  	\caption{\label{Fig1} Wedge propagation in experiment after transition from $q_1=11$\,cm$^2$/s to $q_2=26$\,cm$^2$/s ($W=0.65$\,m, $T=1.2$\,cm). (a) Surface deviation profile at different times ($t$=0, 0.2\,s, 0.75\,s, 1.5\,s, 2.5\,s, 3.5\,s, and 4.6\,s from bottom to top) after change in $q$. The feed zone extends from $x$=0 to the dashed line. Inset: apparatus geometry and wedge propagation mechanism (see text). (b) Temporal evolution of surface height from upstream (largest $h$) to downstream (smallest $h$). Circles indicate the wedge front. (c) Front position vs. time from experiments (circles) and wedge-model prediction (solid curve). Inset: average surface slope $\bar{\alpha}$ vs. $t$ with slope $\bar{\alpha}_1=30.8^\circ$ at $q_1$, wedge angle $\bar{\alpha}_w=32.0^\circ$, and final slope $\bar{\alpha}_2=32.4^\circ$ at $q_2$.}
  	\vspace*{-15 pt}
  \end{figure*}

\par The experimental setup (inset of Fig.~\ref{Fig1}) consists of two parallel vertical rectangular plates - an aluminum back wall and a glass front wall for visualization. The gap between the front and the back plates $T$ and the width of the apparatus $W$ are set by vertical spacers placed between the plates, as described previously~\cite{Fan2012}. Monodisperse non-cohesive glass spheres of diameter $d=1.18\pm0.07~\text{mm}$ and material density $\rho=2440~\text{kg/m}^{3}$ are fed into the rectangular container near the left side from a height at 0.6\,m above the bottom wall by an auger feeder to form a quasi-2D one sided heap. In each experiment, the feed rate $q(t)$ is first set to $q_1$ until the heap extends the entire width of the apparatus and the flow is fully developed~\cite{Fan2012,Fan2013}. Then, the feed rate is changed to a different value $q_2$ in approximately 0.2~s, a negligible duration compared to the transient process duration (several seconds). Neither static electricity nor humidity appear to significantly influence the experimental results. To capture the heap evolution, videos of the entire heap are recorded at 60~frames/s with a spatial resolution of $0.7~\text{mm}$. The free surface of the entire heap $h$ is identified by the sharp transition in the image intensity between the background and particles. The location of the surface is averaged over 5\,mm wide horizontal bins for 0.1\,s (6 frames), and over five identical experiments to reduce uncertainty.

\par An example of a slow-to-fast feed rate transition ($q_2>q_1$) starting at $t=0$ is shown in Fig.~\ref{Fig1}. The coordinate system's origin is at the bottom left corner of the heap with $x$ in the horizontal direction and $z$ in the vertical direction. To demonstrate how the transient surface height trajectory $h(x,t)$ deviates from the $q_1$ steady state trajectory, where the entire surface rises with velocity $v_{r1}=q_1/W$, we plot the surface deviation, $\tilde{h}(x,t)=h(x,t)-h(x,0)-v_{r1}t$, at different times (Fig.~\ref{Fig1}(a)). After the feed rate is increased to $q_2$, the surface near the feed zone rises faster than the downstream portion of the surface forming a wedge of material with an average surface angle $\bar{\alpha}_w$ steeper than the steady state average surface angle $\bar{\alpha}_1$ under $q_1$, while the rest of the surface continues to rise at $v_{r1}$, indicated by $\tilde{h}=0$. As time advances, the wedge grows until its front edge reaches the downstream wall. The propagation of the wedge front is clearer in Fig.~\ref{Fig1}(b), which shows the surface height evolution at equally spaced streamwise locations: the top (bottom) curve corresponds to the furthest upstream (downstream) location. The slope of these curves is the local surface rise velocity, which would be constant if the feed rate were unchanged. However, after the feed rate increases at $t$=0, the surface portion furthest upstream (top curve) responds almost instantaneously and starts to rise faster, resulting in its height deviating from $\tilde{h}=0$ (dashed line), while the downstream portion (lower curves) remains in the state associated with $q=q_1$. The location of the wedge front, defined as the location where $\tilde{h}$ becomes larger than $d$/2, is indicated by circles in Fig.~\ref{Fig1}(b).

 \begin{figure*}
 	\centerline{\includegraphics[width=7.2 in]{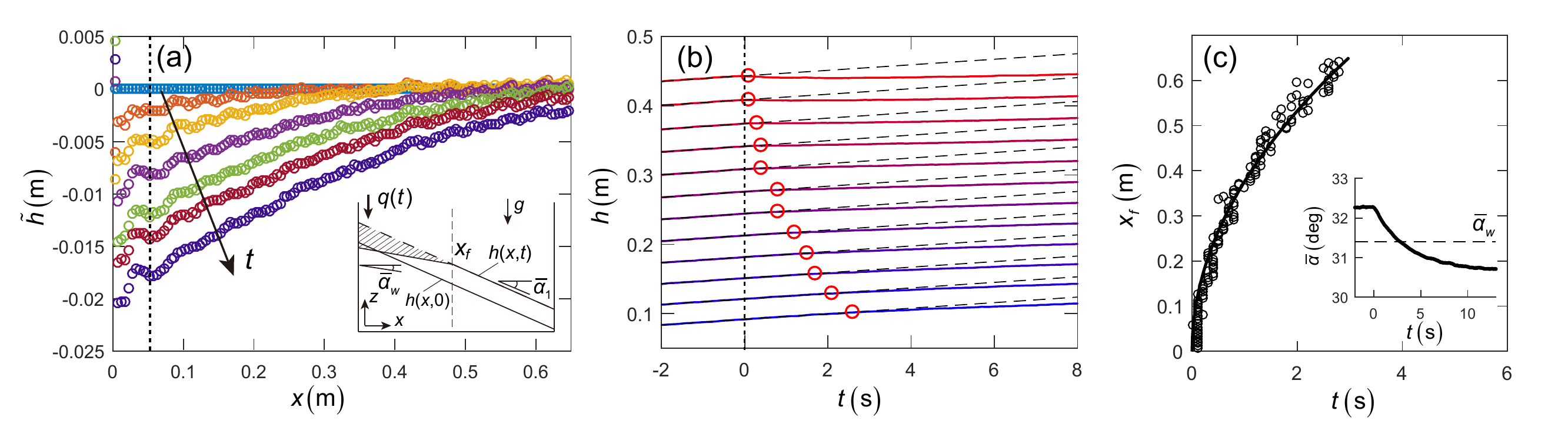}}
 	\vspace*{-10 pt}
 	\caption{\label{Fig2} Wedge propagation details in experiment after transition from $q_1=26$\,cm$^2$/s to $q_2=11$\,cm$^2$/s ($W=0.65$\,m, $T=1.2$\,cm). (a) Surface deviation profile at different times ($t$=0, 0.3\,s, 0.75\,s, 1.5\,s, 2.3\,s, 3.0\,s, and 4.0\,s from bottom to top) after change in $q$. The feed zone extends from $x$=0 to the dashed line. Inset: wedge propagation mechanism (see text). (b) Temporal evolution of surface height from upstream (largest $h$) to downstream (smallest $h$). Circles indicate the wedge front. (c) Front position vs. time from experiment (circles) and model prediction (solid curve). Inset: average surface slope $\bar{\alpha}$ vs. $t$ with slope $\bar{\alpha}_1=32.3^\circ$ at $q_1$, wedge angle $\bar{\alpha}_w=31.4^\circ$, and final slope $\bar{\alpha}_2=30.8^\circ$ at $q_2$.}
 	\vspace*{-15 pt}
 \end{figure*}

\par The transient process described above differs from those in tumblers~\cite{Pohlman2009,CourrechDuPont2005} and streamwise invariant flows~\cite{Jop2007,parez2016unsteady,taberlet2004growth,Capart2015} where the entire flowing layer responds to the external change (e.g., in different rotation speed or tilt) instantaneously. Here, the transient process originates from the feed zone, and its influence propagates gradually downstream in the form of a wedge front. The front slows as it travels downstream, see Fig.~\ref{Fig1}(b), which distinguishes it from avalanche fronts with constant propagation velocities observed~\cite{Douady2002,makse1998dynamics,borzsonyi2005two,edwards2015erosion,saingier2016front} or assumed~\cite{Gray2010,Boutreux1998} in previous studies. This slowing indicates a different driving mechanism. To model the evolution of free surface granular flows, depth integrated continuity and momentum equations have been applied using various constitutive laws~\cite{Douady2002,borzsonyi2005two,edwards2015erosion,saingier2016front}. However, for bounded heap flows, special treatment of the erosion and deposition between the flowing layer and the static bed is necessary~\cite{Douady2002,Capart2015,jenkins2016erosion}, and both the time development and the streamwise gradient have to be resolved, which substantially complicates the problem. Alternatively, a simple but effective approach based on geometric arguments describes the wedge development. By integrating $\tilde{h}$ in the $x$ direction, an overall mass conservation equation is obtained:

\begin{equation}
\int\limits_{0}^{W}{\tilde{h}dx}=\int\limits_{0}^{W}{\left[ h(x,t)-h(x,0)\right] dx}-\int\limits_{0}^{W}{v_{r1}tdx}.
\label{Eq1} 
\vspace*{-0 pt}
\end{equation}

\noindent The l.h.s.\ of Eq.~(\ref{Eq1}) is the area of the growing wedge shown in the inset of Fig.~\ref{Fig1}(a), which can be approximated by the area of a triangle, $\frac{1}{2}x_f^2(\tan{\bar{\alpha}_w}-\tan{\bar{\alpha}_1})$, where $x_f$ is the instantaneous front position. The wedge angle $\bar{\alpha}_w$, which only varies slightly as the wedge front propagates, is measured as the mean heap surface angle $\bar{\alpha}$ at the point when the wedge reaches the endwall. The first term on the r.h.s.\ is the increase of the heap area, $q_2t$, while the second term is the increase in heap area had the feed rate been maintained at $q_1$, namely $q_1t$. Substituting these expressions into Eq.~(\ref{Eq1}) gives an approximation for the instantaneous front position,

\begin{equation}
x_f=\sqrt{Ct},
\label{Eq2} 
\vspace*{-5 pt}
\end{equation}

\noindent where $C=\frac{2(q_2-q_1)}{ \tan{\bar{\alpha}_w}-\tan{\bar{\alpha}_1}}$ is a propagation constant dependent only on parameters of the problem. Eq.~\ref{Eq2} agrees well with the experimentally measured front position (Fig.~\ref{Fig1}(c)), which implies that this transient process can be viewed as filling an additional wedge on top of a rising heap surface with the front propagation velocity decreasing as $1/\sqrt{t}$ as the wedge grows. 

\par For the fast-to-slow feed rate transition ($q_2<q_1$), the physics is similar. After $q$ is reduced to $q_2$, the rise velocity of the surface near the feed zone decreases, resulting in $\tilde{h}<0$ in the upstream portion of the flow, and a ``negative" wedge propagating downstream (Fig.~\ref{Fig2}(a)) until it reaches the endwall. Figure.~\ref{Fig2}(b) shows the gradual deviation of the surface from the previous state (dashed line) from upstream to downstream, and a slowing front, similar to the slow-to-fast transition. Due to this similarity~\footnote{The 0.1$^\circ$ difference between $\bar{\alpha}$ at the high feed rate noted in the captions of Figs.~\ref{Fig1} and~\ref{Fig2} occurs because of different distances ($\approx$5\,cm) between the feeder and the heap surface~\cite{grasselli1999shapes} when $\bar{\alpha}$ was measured in the two cases.}, Eq.~(\ref{Eq2}) can be directly applied to this type of front propagation, but with negative $q_2-q_1$ and $\tan{\bar{\alpha}_w}-\tan{\bar{\alpha}_1}$, and it again accurately predicts the observed $\sqrt{t}$ advance of the wedge front, see Fig.~\ref{Fig2}(c). 

\par Although the wedge approximation is suitable for predicting front propagation for both increasing and decreasing feed rate, it is clear that the $\bar{h}$ profile in the upstream region of the front is slightly curved (Figs.~\ref{Fig1}(a) and~\ref{Fig2}(a)). Moreover, since the final slope $\bar{\alpha}_2 \ne \bar{\alpha}_w$ (inset of Figs.\ref{Fig1}(c) and~\ref{Fig2}(c)), an additional transient process exists after the wedge reaches the endwall, indicating that additional physics is needed to more accurately describe the transient response. 

\par To explore the underlying physics, additional kinematic details are extracted from DEM simulations of single transitions using an in-house code that was previously applied and validated in heaps~\cite{Fan2013,Xiao2016,fan2015shear}. Particle interactions are modeled with a linear spring-dashpot normal force and a combination of linear spring and Coulomb friction tangential force~\cite{Fan2013}. To reduce computation cost, we simulate the system using slightly larger $d=2\pm0.2$\,mm particles with restitution coefficient $e=$0.8, particle-particle and particle-wall friction $\mu=$0.4, and a binary collision time of $t_c=$1$\times10^{-4}$\,s~\cite{Fan2013}. The feed position is kept at a constant height ($\approx$20\,cm) above the left end of heap surface to eliminate the influence of changing drop height~\cite{grasselli1999shapes}. The integration time step is $t_c/40$ for numerical stability~\cite{Fan2013}. Instantaneous horizontal flux profiles $q(x,t)$ and local surface slope $\partial{h(x,t)}/\partial{x}$ are calculated from the simulation results. To determine $q(x,t)$, the horizontal velocity $u(x,z)$ is computed using a volumetric binning method with bin size 10\,mm$\times$2\,mm~\cite{Fan2013}. To further reduce uncertainty, data are averaged over 0.05\,s and over 5 repeated simulations. The instantaneous horizontal flux profiles are calculated as $q(x)=\frac{1}{\bar{\phi}}\int\limits_{0}^{h}u\phi dz$, where $\phi$ is the local packing fraction and $\bar{\phi}=0.58$ is the packing fraction averaged over the entire heap. The instantaneous local slopes are calculated by fitting a line to $h$ over a 5\,cm interval at each horizontal location corresponding to a bin center. Three feed rates are considered (11\,cm$^2$/s, 35\,cm$^2$/s, and 69\,cm$^2$/s), and the single transitions between these feed rates as well as steady flows are simulated. 

\par
\begin{figure}
 	\centerline{\includegraphics[width=3.0 in]{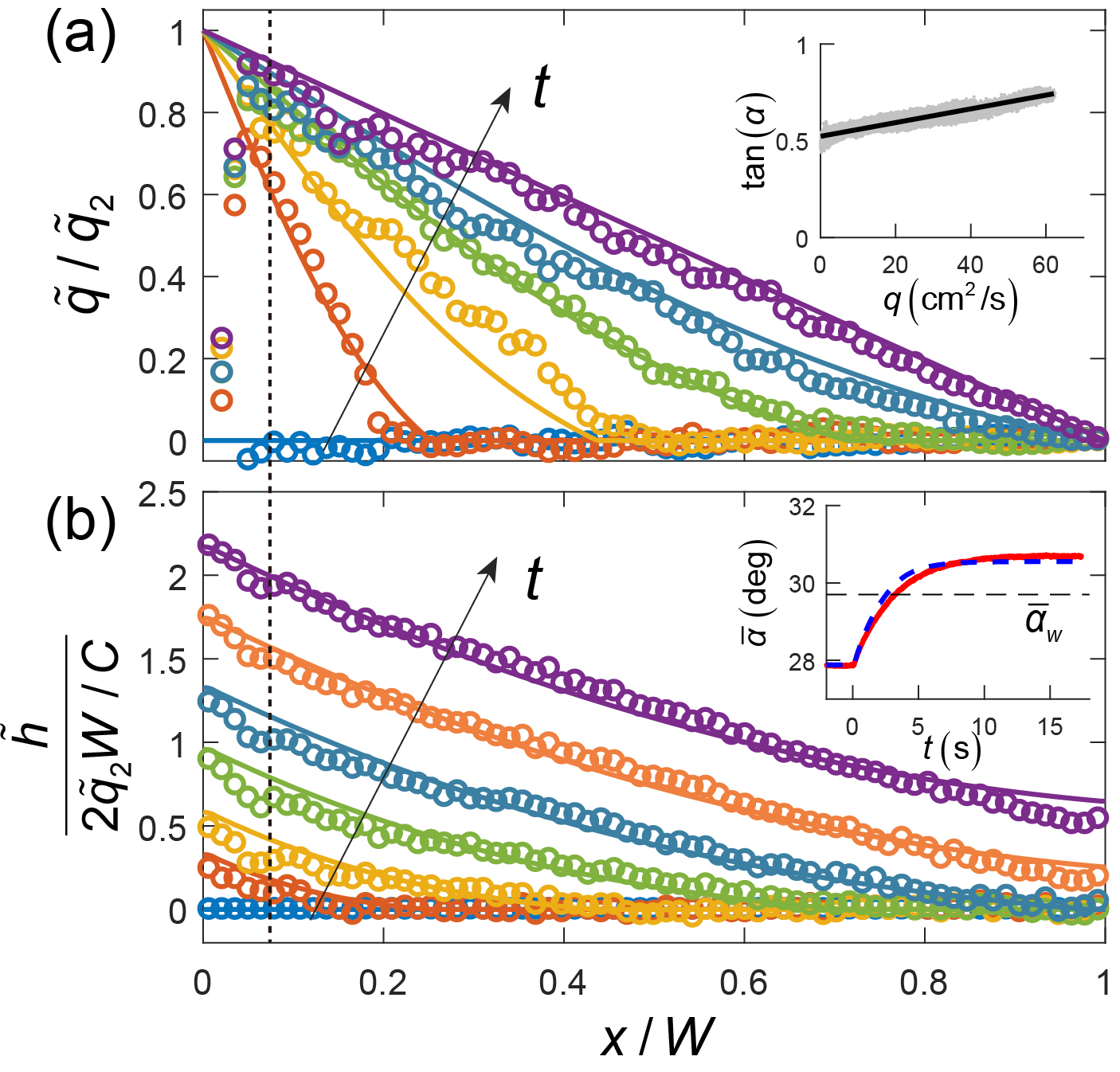}}
 	\vspace*{-10 pt}
 	\caption{\label{Fig3}DEM simulation results (circles) and model results (solid curves) for a single transition from $q_1=11$\,cm$^2$/s to $q_2=35$\,cm$^2$/s in a system with $W$=0.69\,m and $T$=12\,cm. $\bar{\alpha}_1=27.9^\circ$ and $\bar{\alpha}_w=29.7^\circ$. The feed zone extends from $x=0$ to the dashed line. (a) $\tilde{q}(x,t)$ profiles at $t$=0, 0.25\,s, 0.75\,s, 2.0\,s, 3.8\,s, and 9.8\,s from bottom to top. Inset: $\tan\alpha$ vs. $q$ from steady flow and single transition measurements (gray dots) with solid line $q=-A\partial{h}/\partial{x}+B$, where $A$=0.028\,m$^2$/s and $B=-0.015$\,m$^2$/s. (b) $\tilde{h}(x,t)$ profiles at $t$=0, 0.25\,s, 0.75\,s, 2.0\,s, 3.8\,s, 6.5\,s, and 9.8\,s from bottom to top. Inset: $\bar{\alpha}$ vs. $t$ from simulation (red curve) and our model (blue dashed curve). }
 	\vspace*{-15 pt}
 \end{figure}

 To better model the transient surface dynamics, we first quantify the relationship between the local slope, $\partial{h}/\partial{x}$ and the local flow rate $q(x)$. The inset in Fig.~\ref{Fig3} shows local instantaneous measurements of $\partial{h}/\partial{x}$ vs. $q$ for both steady flows and single transitions ($\approx$85000 data points). Note that $\partial{h}/\partial{x}=-\tan{\alpha}$, where $\alpha$ is the local surface angle. The data for both steady and transient flows collapse indicating that within the range of flow rates simulated, the relation between the local slope and the local flow rate is universal, and unsteadiness (i.e.\ $\partial{q}/\partial{t}\neq0$) plays only a minor role as evidenced by the scatter due to a small hysteresis between increasing and decreasing $q$. This relation can be approximated as $q=-A\partial{h}/\partial{x}+B$, where $A$ and $B$ are constants. Similar to $\tilde{h}$, we introduce the flux deviation, $\tilde{q}(x,t)=q(x,t)-q_1(1-x/W)$, as the deviation of the instantaneous local flux from the steady state value noting that the flow rate under $q_1$ decreases linearly with horizontal position~\cite{Fan2013}. Substituting $\tilde{q}$ and $\tilde{h}$ into the relation between $q$ and $\partial{h}/\partial{x}$ gives,

\begin{equation}
\tilde{q}=-A\frac{\partial{\tilde{h}}}{\partial{x}}.
\label{Eq3}
\vspace*{-0 pt}
\end{equation}

\noindent Similarly, using expressions for $\tilde{q}$ and $\tilde{h}$, it can be shown that the continuity equation~\cite{Douady2002}, $\partial{h}/\partial{t}+\partial{q}/\partial{x}=0$, can be expressed as 

\begin{equation}
\frac{\partial{\tilde{h}}}{\partial t}+\frac{\partial {\tilde{q}}}{\partial x}=0.
\label{Eq4}
\vspace*{-0 pt}
\end{equation}

\noindent Note that Eq.~\ref{Eq3} takes the form of Fick's law with $\tilde{h}$ in place of concentration, $\tilde{q}$ in place of the diffusion flux, and $A$ in place of the diffusion coefficient. Eq.~\ref{Eq3} can be used to express Eq.~\ref{Eq4} in terms of $\tilde{h}$, i.e.,\ $\partial \tilde{h}/\partial t=A\partial ^2 \tilde{h}/\partial x^2$, which results in an equation having the same form as the diffusion equation.  Equivalently, differentiating Eq.~\ref{Eq3} with respect to $t$ and Eq.~\ref{Eq4} with respect to $x$ and then combining, gives

\begin{equation}
\frac{\partial{\tilde{q}}}{\partial t}=A \frac{\partial^2{\tilde{q}}}{\partial x^2}.
\label{Eq5}
\vspace*{-0 pt}
\end{equation}

\noindent In the bounded heap, the upstream boundary condition is $\tilde{q}(0,t)=\tilde{q}_2$, where $\tilde{q}_2=q_2-q_1$. Before the wedge reaches the endwall, Eq.~\ref{Eq5} applies only in the wedge which defines a downstream moving boundary condition of $\tilde{q}(x_f,t)=0$, where $x_f$ is given by Eq.~\ref{Eq2}. Since $x_f$ increases as $\sqrt{t}$, a similarity solution can be obtained~\cite{hu1996mathematical} by choosing a similarity variable $\xi=x/\sqrt{t}$:

\begin{equation}
\tilde{q}(x,t)=\tilde{q}_2\left[1-\frac{\text{erf}\left(\frac{x}{\sqrt{4tA}}\right)}{\text{erf}\left(\frac{x_f}{\sqrt{4tA}}\right)}\right].
\label{Eq6}
\vspace*{-0 pt}
\end{equation}

\noindent For the region ahead of the front ($x_f < x \le W$), $\tilde{q}(x,t)=0$. After the wedge reaches the endwall, the right boundary condition becomes $\tilde{q}(W,t)=0$, and transient solutions are obtained numerically by a standard implicit finite difference method. $\tilde{h}(x,t)$ is determined numerically by integrating Eq.~\ref{Eq4}.

\par Examples of $\tilde{q}$ profiles (normalized by $\tilde{q}_2$) and $\tilde{h}$ profiles (normalized by $2\tilde{q}_2W/C=W(\tan{\bar{\alpha}_w}-\tan{\bar{\alpha}_1})$) using the above formalism in a slow-to-fast transition are shown in Fig.~\ref{Fig3}. The analytic approach agrees well with the corresponding DEM simulation. Near the beginning of the transition ($t$=0.25\,s), $\tilde{q}$ increases sharply near the feed zone (Fig.~\ref{Fig3}(a)) and the strong gradient results in increased local deposition of particles on the heap leading to the formation of the wedge (Fig.~\ref{Fig3}(b)). As the change in $\tilde{q}$ further propagates downstream ($t$=0.75\,s and 2.0\,s), the transition at the front becomes smoother, and both the $\tilde{q}$ and $\tilde{h}$ profiles for $x \le x_f$ become slightly curved because of Eq.~\ref{Eq5}. Here, a smaller $A$ results in larger profile curvature, much like diffusion with a smaller diffusivity. 

\par After the front reaches the endwall ($t$=3.8\,s), the curved $\tilde{q}$ profile continues to evolve towards the new linear steady state ($t$=9.8\,s). This slow evolution after the front reaches the endwall corresponds to the additional change in $\bar{\alpha}$ shown in the insets of Fig.~\ref{Fig1}(c) and~\ref{Fig2}(c), and is well captured by the model (Fig.~\ref{Fig3}(b) inset).  Again, smaller $A$ results in a longer transition duration, much like a substance with smaller diffusivity. Since the propagation constant $C$ depends on average surface slope and feed rate, a relation between $C$ and $A$ likely exist, similar to that in other moving boundary problems~\cite{hu1996mathematical,krapivsky1996life,ozisik1993heat}. Moreover, as $A$, $B$, and $C$ are dimensional constants, scalings likely exist between these constants and physical parameters such as the particle diameter and the flowing layer depth. The small differences between the model and simulation results are likely due to approximating the relation between $\alpha$ and $q$ as linear and neglecting $\partial q /\partial t$ and higher order spatial derivatives in Eq.~\ref{Eq3}. We have obtained similar quantitative agreement between the model and DEM simulation results for all combinations of the three feed rates including both fast-to-slow and slow-to-fast transitions.

\par The model developed here for transient granular flow in a bounded heap under a step change in the feed rate captures the heap transition dynamics observed in experiments and simulations, and can potentially be applied to other depositing flows such as open heap flows and tumbler flows. Note, however, that Eq.~\ref{Eq5} depends on a linear relationship between the local slope and flux, a relationship that is apparently due to the frictional interaction of the flowing grains with the sidewalls in a narrow gap geometry~\cite{jop2005crucial,taberlet2003superstable,midi2004dense}.
For transient heap flows with wider gaps where sidewall friction has a smaller influence or 3D conical heap flows, this relation may be non-linear~\cite{jop2005crucial}, resulting in a different form for Eq.~\ref{Eq5}, which may produce different transients.

\par Funded by NSF Grant \# CBET-1511450.

\end{document}